\documentclass[a4paper,fleqn]{article} 
 \usepackage{latexsym}
 \usepackage{doc}
 \usepackage{amsfonts}
 \usepackage{exscale}
 \usepackage{fontenc}
 
\begin{document}
 \newcommand{\nw}{\newcommand}
 \nw{\be}{\begin{equation}}
 \nw{\ee}{\end{equation}}
 \nw{\hs}{\hspace}
 \nw{\vs}{\vspace}
 \nw{\nt}{\noindent}
 \nw{\qd}{\quad}
 \nw{\qq}{\qquad}
 \nw{\bd}{\begin{displaymath}}
 \nw{\ed}{\end{displaymath}}
 \nw{\ba}{\begin{array}}
 \nw{\ea}{\end{array}}
 \nw{\re}{\renewcommand}
 \nw{\lt}{\left}
 \nw{\rt}{\right}
 \nw{\bq}{\begin{eqnarray}}
 \nw{\eq}{\end{eqnarray}}
 \nw{\nn}{\nonumber}
 \nw{\cd}{\cdots}
 \nw{\ld}{\ldots}
 \nw{\lb}{\linebreak}
 \nw{\mb}{\mbox}
 \nw{\bm}{\bibitem}
 \nw{\ce}{\cite}
 \title{A unified treatment of exactly solvable and quasi-exactly solvable quantum potentials}
 \author{B. Bagchi\footnote{bbagchi@cucc.ernet.in} and 
                       A. Ganguly\footnote{asish@cucc.ernet.in, gangulyasish@rediffmail.com} \\
    Department of Applied Mathematics, University of Calcutta, \\ 92 Acharya Prafulla Chandra Road, Kolkata -- 700009, India} 
 \date{17 January 2003}
 \maketitle
 \begin{abstract}
 By exploiting the hidden algebraic structure of the Schr\"{o}dinger Hamiltonian, namely the sl(2), we 
 propose a unified approach of generating both exactly solvable and quasi-exactly solvable potentials. We
 obtain, in this way, two new classes of quasi-exactly solvable systems one of which is of periodic type
 while the other hyperbolic.

 \vs{2mm}
 \nt
 \hs{-.8cm}PACS number(s): 03.65
 \end{abstract}

 Tracking down solvable quantum potentials has always aroused interest. Apart from being useful in the 
 understanding of many physical phenomena, the importance of searching for them also stems from the fact
 that they very often provide a good starting point for undertaking perturbative calculations of more
 complex systems.

 Solvable potentials can be broadly classified into two categories : the ones which are exactly 
 solvable\ce{bha,nat,gin,dab}(including the conditional ones\ce{dut,raj}) and others which are quasi-exactly 
 solvable\ce{turb,shif,uly,ush}. A spectral problem is said to be exactly solvable(ES) if one can determine 
 the whole spectrum analytically by a finite number of algebraic steps. Factorization 
 hypothesis\ce{inf,sam}, group-theoretical techniques with a spectrum-generating algebra\ce{cord,wu,bb1} 
 and use of integral transformations\ce{abh,pur} are some of the time-honoured procedures of constructing ES  
 potentials\ce{bbbook}. On the other hand, there exist an infinite number of normal spectral 
 problems which are not amenable to an exact treatment. These are the non-solvable (NS) ones. The 
 quasi-exactly solvable(QES) class is the missing link\ce{as3,as4as5} between the ES and the NS potentials. 
 Actually for a QES system we can only determine a part of the whole spectrum : this essentially means that 
 in an infinite-dimensional space of states there exists a finite-dimensional subspace for which the 
 Schr\"{o}dinger equation admits partial algebraization.

 However, in the literature, a common framework that brings together the ES and QES class is still lacking. 
 The purpose of this letter is to fill this gap by exploiting the hidden 
 dynamical symmetry of the Schr\"{o}dinger equation. We show, in a straightforward way, that by subjecting
 the Schr\"{o}dinger equation to some coordinate transformation and adopting for the underlying symmetry
 group the simplest choice namely the sl(2), it is possible to set up a master equation from which the
 ES and QES potentials readily follow. Following this line, we construct not only some of the well-known ES 
 potentials which are with us for a long time but also uncover new families of QES potentials which 
 hitherto have remained unnoticed.

 Let us start with the following differential realization of the sl(2) generators $T^{\pm},T^{0}$ given by
 \be
 \hs{1cm}T^{+}=\xi^{2}\frac{d}{d\xi}-n\xi \, , \qd T^{-}=\frac{d}{d\xi} \, , \qd 
                                  T^{0}=\xi\frac{d}{d\xi}-\frac{1}{2}n \, , 
 \ee
 where $\xi \in \mathbb{R}$ and $n$ is some non-negative integer. These generators act on the represention 
 space $\mathcal{P}_{n}$ of polynomials in $\xi$ of degree not exceeding $n$ and obey the commutation 
 relations
 \be
 \hs{2cm} [T^{+},T^{-}]=-2T^{0} \, , \qq [T^{0},T^{\pm}]=\pm T^{\pm} \, .
 \ee

 Let us assume that the quantum Hamiltonian is expressible as a quadratic combinations of the generators
 $T^{a}$ with constant coefficients, that is
 \be
 \hs{2cm} \mathcal{H}=-\sum_{a,b=0,\pm}C_{ab}T^{a}T^{b}-\sum_{a=0,\pm}C_{a}T^{a}-d(n) \, ,
 \ee
 where $C_{ab},C_{a}$ are numerical parameters of which $C_{ab}$ is symmetric and $d$ is a suitably chosen 
 constant that depends on $n$. Note that this functional dependence is single-valued for ES models, while it 
 is multivalued for QES. In the latter case the range is $\{d_{0}(n),d_{1}(n),\ld d_{n}(n)\}$.

 Now from (1) it is easy to see that $\mathcal{H}$ has the representation
 \be
 \hs{4cm} \mathcal{H}(\xi)=-\sum^{4}_{j=2}B_{j}(\xi)\frac{d^{j-2}}{d\xi^{j-2}} \, .
 \ee
 The coefficients $B_{j}$'s in (4) are the $j$-th degree polynomial in $\xi$ :
 \bq
 B_{4}(\xi) & = & C_{++}\xi^{4}+2C_{+0}\xi^{3}+C_{00}\xi^{2}+2C_{0-}\xi+C_{--} \, , \\ \nn 
 B_{3}(\xi) & = & \frac{1-n}{2}\frac{dB_{4}}{d\xi}+A_{2}(\xi) \, , \\
 B_{2}(\xi) & = & \frac{n(n-1)}{12}\frac{d^{2}B_{4}}{d\xi^{2}}-\frac{n}{2}\frac{dA_{2}}{d\xi}
                       +\frac{n(n+2)}{12}C_{00}+d(n) \, , \nn
 \eq
 where $C_{+-}=C_{-+}=0$ because of the constancy of the Casimir operator and
 $A_{2}(\xi)=C_{+}\xi^{2}+C_{0}\xi+C_{-} \, .$

 However the coeffecient of $d^{2}/d\xi^{2}$ in (4) is not unity. To achieve this we introduce the mapping
 $u(\xi)=\int^{\xi} \, [B_{4}(\tau)]^{-1/2}d\tau$ thereby obtaining
 \be
 \mathcal{H}(u)=\lt .-\frac{d^{2}}{du^{2}}+\lt .\frac{1}{2\sqrt{B_{4}}}\lt (\frac{dB_{4}}{d\xi}-2B_{3}\rt )
             \rt |_{\xi=\xi(u)}\frac{d}{du}-B_{2} \rt |_{\xi=\xi(u)} \, .
 \ee

 To proceed further, we may look upon $\mathcal{H}(u)$ as a `coordinate-transformed' Schr\"{o}dinger
 Hamiltonian. Indeed let us consider a change of variable $x\rightarrow x(u)$ that transforms the
 Schr\"{o}dinger wave function according to
 \be
 \hs{4cm} \psi(x)\rightarrow g(x)\chi(u(x)) \, .
 \ee
 The standard Schr\"{o}dinger equation (with $\hbar=2m=1$) with the potential $V(x)$
 \be
 \hs{2cm} \lt [ -\frac{d^{2}}{dx^{2}}+V(x)\rt ]\psi(x)=E\psi(x)
 \ee
 is then recast to the form
 \be
 \hs{2cm} -\frac{d^{2}\chi}{du^{2}}-\lt [ \frac{u''}{u'^{2}}+\frac{2g'}{u'g}\rt ]\frac{d\chi}{du}
                  -\lt [ \frac{g''}{gu'^{2}}+\frac{E-V(x)}{u'^{2}}\rt ]\chi(u)=0 \, ,
 \ee
 where the primes denote derivatives with respect to $x$.

 The general nature of (9) enables one to touch upon those differential equations whose solutions are 
 well-known. In particular those which are associated with special functions can be considered for the 
 determination of solvable potentials. Here we take a different route by seeking a direct
 correspondance of (9) with the `coordinate-transformed' Schr\"{o}dinger Hamiltonian $\mathcal{H}$. We 
 thus obtain 
 \begin{eqnarray}
 \lefteqn{\hs{-.9cm}V(x)=\lt [ E-\frac{u'''}{2u'}+\frac{3}{4}\lt ( \frac{u''}{u'}\rt )^{2} \rt.} \nn \\
   & & \hs{-.6cm}\mb{} \lt. -u'^{2}\lt \{ B_{2}-\frac{1}{4}\lt ( 2\frac{dB_{3}}{d\xi}-
 \frac{d^{2}B_{4}}{d\xi^{2}}\rt )-\frac{1}{16B_{4}}(2B_{3}-\frac{dB_{4}}{d\xi})(2B_{3}-3\frac{dB_{4}}{d\xi})
                           \rt \}_{\xi=\xi(u)}\rt ]_{u=u(x)}
 \end{eqnarray}
 In this connection we may emphasize that whereas for ES models the energy levels form an infinite sequence,
 in the QES case there can be at most ($n+1$) levels for each choice of $n$. It is thus appropriate to label
 the energy levels by an index $j$ depending on $n$ i.~e.\ $j=j(n)$ that runs with $n$ and takes values
 $0,1,\ld ,n,\ld \infty$ for the ES models but assumes only a finite number of values $\{0,1,\ld ,n\}$ for
 QES. 

 In (10) the $B_{j} \, (j=2,3,4)$ functions of sl(2) have to be suitably adjusted against the arbitrary 
 function $u(x)$ of the coordinate transformation to have an acceptable form of the quantum potentials. For
 this the normalizability of the wave function is to be ensured. Note that in arriving at the form (10)
 we have eliminated the function $g(x)$. However, for a particular choice of $u(x)$, it can be determined 
 from
 \be
 \hs{2cm}g(x)=(u')^{-1/2}\exp \lt [\frac{1}{2}\int^{u(x)}\lt \{ \frac{2B_{3}-dB_{4}/d\xi}
                {2\sqrt{B_{4}}} \rt \}_{\xi=\xi(u)}du \rt ] \, .
 \ee
 Knowing $u(x)$ and $g(x)$ the wave function can be found from (7).

 Equation (10) is the central result of our paper : it opens up a new approach of generating ES and QES 
 potentials.

 To see how our scheme works in practice, let us first address to the problem of deriving some well-known
 ES potentials from (10). Without giving the details of our calculations which are straightforword we present
 the results in standard forms\ce{lev} :
 \begin{description}
  \item[$\bullet$ Harmonic oscillator:] $u=x, \, B_{4}=1, \, B_{3}=-\omega \xi, \, B_{2}=n\omega$ [i.\ e.\ 
 we choose $C_{--}=1, \, d(n)=n\omega/2, \, C_{0}=-\omega \, (\omega >0), \, 
 C_{++}=C_{+0}=C_{00}=C_{0-}=C_{+}=C_{-}=0$.]
 \bd
 \hs{2cm}V(x)=\frac{1}{4}\omega^{2}x^{2}, \qq E_{j}=(j+\frac{1}{2})\omega \, ,
 \ed 
 \bd
 \psi_{j}(x)=\mathcal{N}_{j}\exp (-\frac{1}{4}\omega x^{2})H_{j}(\sqrt{\frac{\omega}{2}}x), \qd 
                          j=0,1,\ld ,n,\ld ,\infty \, ,
 \ed
 $\mathcal{N}_{j}$ being the normalization constant.
 \item[$\bullet$ Morse:] $u=x, \, B_{4}=\alpha^{2}\xi^{2}, \, B_{3}=\alpha(\alpha-2A)\xi+2B\alpha, \,
  B_{2}=-n^{2}\alpha^{2}+2An\alpha$ [i.\ e.\ $C_{00}=\alpha^{2}, \, C_{++}=C_{+0}=C_{0-}=C_{--}=C_{+}=0, \,
  C_{0}=\alpha(n\alpha-2A), \, d(n)=An\alpha-3n^{2}\alpha^{2}/4, \, C_{-}=2B\alpha$.] 
 \bd
  V(x)=B^{2}\exp [-2\alpha x]-B(2A+\alpha)\exp [-\alpha x], \qd E_{j}=-(A-j\alpha)^{2}, 
 \ed
 \bd
 \hs{-1cm}\psi_{j}(x)=\mathcal{N}_{j}\exp [(j\alpha-A)x]
                                \exp [-\frac{B}{\alpha}e^{-\alpha x}]L^{(2\frac{A}{\alpha}
 -2j)}_{j}\lt ( \frac{2B}{\alpha}e^{-\alpha x} \rt ), \qd j=0,1,\ld ,n,\ld \infty \, ,
 \ed
 $\mathcal{N}_{j}$ being the normalization constant.
 \item[$\bullet$ P\"{o}schl-Teller:] $u=x, \, B_{4}=4\alpha^{2}(\xi^{2}-1), \, 
 B_{3}=4\alpha\{(A+B+2\alpha)\xi+B-A-\alpha\}, \, 
  B_{2}=\alpha^{2}(1-4n^{2})+2\alpha \{ 2n(A-B)+A+B \}+4AB$ [i.\ e.\ $C_{00}=4\alpha^{2}=-C_{--}, \, 
 C_{++}=C_{+0}=C_{0-}=C_{+}=0, \, d(n)=4AB+\alpha^{2}(1+3n)(1-n)+2n\alpha(3A-B)+2\alpha(A+B), \, 
 C_{-}=4\alpha(B-A-\alpha), \, C_{0}=4\alpha\{A+B+\alpha(n+1)\}$.] 
 \bd
 V(x)=B(B-\alpha)\mb{cosech}^{2}\alpha x-A(A+\alpha)\mb{sech}^{2}\alpha x, \qd E_{j}=-(A-B-2j\alpha)^{2},
 \ed
 \bd
 \hs{-1cm}\psi_{j}(x)=\mathcal{N}_{j}\sinh^{B/ \alpha}\alpha x\cosh^{-A/ \alpha}\alpha x \, P_{j}%
 ^{(\frac{B}{\alpha}-\frac{1}{2},-\frac{A}{\alpha}-\frac{1}{2})}(\cosh 2\alpha x), \, 
                      j=0,1,\ld ,n,\ld \infty \, ,
 \ed
 $\mathcal{N}_{j}$ being the normalization constant.
 \item[$\bullet$ Scarf~II:] $u=x, \, B_{4}=\alpha^{2}(\xi^{2}+1), \, B_{3}=\alpha(2A+3\alpha)\xi
 +2B\alpha, \, B_{2}=\alpha(n+1)\{\alpha(1-n)+2A\}$[i.\ e.\ we choose $C_{00}=C_{--}=\alpha^{2}, \, 
 C_{0}=\alpha^{2}(n+2)+2A\alpha, \, C_{-}=2B\alpha, \, C_{++}=C_{+0}=C_{0-}=C_{+}=0, \, 
 d(n)=\alpha^{2}+A\alpha(3n+2)+n\alpha^{2}(4-3n)/4$.] 
 \bd
 V(x)=[B^{2}-A(A+\alpha)]\mb{sech}^{2}\alpha x+B(2A+\alpha)\mb{sech}\alpha x \, \tanh \alpha x, 
 \ed
 \bd 
 \hs{3cm} E_{j}=-(A-j\alpha)^{2}, \qd j=0,1,\ld ,n, \ld \infty \, ,
 \ed 
 \bd
 \hs{-1cm}\psi_{j}(x)=\mathcal{N}_{j}\cosh^{-A/\alpha}\alpha x
                                  \exp [-\frac{B}{\alpha}\tan^{-1}(\sinh \alpha x)]
 P_{j}^{(-i\frac{B}{\alpha}-\frac{A}{\alpha}-\frac{1}{2},i\frac{B}{\alpha}-\frac{A}{\alpha}-\frac{1}{2})}
                                                              (i\sinh \alpha x) \, ,
 \ed
 $\mathcal{N}_{j}$ being the normalization constant.
 \item[$\bullet$ Coulomb:] $u=2\sqrt{x} \, (x>0), \: B_{4}=4\xi, \, B_{3}=e^{2}\xi/(n+l+1)+8(l+1), \,
  B_{2}=e^{2}(n+2l+2)/(n+l+1)$ [i.\ e.\ $C_{0-}=2, \, C_{0}=e^{2}/(n+l+1) \, (l\geq 0), \, C_{-}=2(4l+n+3), \,
 C_{++}=C_{+0}=C_{00}=C_{--}=C_{+}=0, \, d(n)=e^{2}(3n+4l+4)/2(n+l+1)$.]
 \bd
 V(x)=-\frac{e^{2}}{x}+\frac{l(l+1)}{x^{2}} \, (0<x<\infty), \qd E_{j}=-\frac{e^{4}}{4(j+l+1)^{2}},
 \ed
 \bd
 \hs{-1cm}\psi_{j}(x)=\mathcal{N}_{j} \, x^{l+1}\exp \lt [ -\frac{e^{2}x}{2(n+l+1)}\rt ] 
                L_{n}^{(2l+1)}\lt ( \frac{e^{2}x}{n+l+1}\rt ), \, j=0,1,\ld ,n,\ld \infty \, ,
 \ed
 $\mathcal{N}_{j}$ being the normalization constant.
 \end{description}

 Having dealt successfully with the generation of ES potentials all of which have been well studied in the
 literature, let us turn to the problem of finding QES potentials from (10). In the following we present two
 new families of QES potentials, one of which is periodic while the other is hyperbolic.
 \section*{A. Periodic model}
 
 Consider the transformation $u=x-a$ ($a\in \mathbb{R}$) along with the representation  
 $B_{4}=\beta^{2}(1-\xi^{2})$ [this comes about if we set $C_{++}=C_{+0}=C_{0-}=0, \,
  C_{00}=-\beta^{2}=-C_{--},\beta\neq 0$]. This particular form for $B_{4}$ immediately yields
 $\xi=\cos \beta u$ and facilitates generating a periodic QES system. Indeed, trialing with various 
 choices of $B_{3}$ and $B_{2}$ we have found that in all four algebraizations exist each leading to
 a distinct QES family. Our results are
 \begin{enumerate}
 \item  $B_{3}=-\alpha \xi^{2}-2\beta^{2}\xi+\alpha\pm\beta^{2}, \, B_{2}=n\alpha \xi+\frac{n(n+2)}{4}
                                                                 \beta^{2}+d_{j}(n)$
 \bd
 \hs{-1cm}\mb{[ i.\ e.\ }C_{+}=-\alpha=\pm\beta^{2}-C_{-}, \, C_{0}=-(n+1)\beta^{2}, 
                            \alpha \neq 0\mb{ ]}
 \ed
 \bd
   V_{1}(x)=-\frac{\alpha^{2}}{8\beta^{2}}\cos 2\beta(x-a)-\alpha (n+1)\cos \beta (x-a)-
                                              \frac{\beta^{2}}{4}
 \ed
 \bd
 \hs{1cm}E_{j}=\frac{n(n+2)}{4}\beta^{2}-\frac{\alpha^{2}}{8\beta^{2}}\mp\frac{\alpha}{2}+d_{j}(n), \qd
                                   j=0,1,\ld ,n \, ,
 \ed
 \bd
 \hs{-1.3cm}\hs{-7pt}\psi_{j}(x)=\sin \lt \{ (\delta_{k+})\frac{\pi}{2}+\beta \frac{x-a}{2}\rt \}
    \exp [-\frac{\alpha}{\beta^{2}}\sin^{2}(\beta \frac{x-a}{2})]
     \sum_{r=0}^{n}b_{j}^{(r)}\cos^{r}\beta (x-a),  (k=+,-).
 \ed
  \item $B_{3}=\alpha \xi^{2}-2\beta^{2}\xi-\alpha\pm\beta^{2}, \, 
                                  B_{2}=-n\alpha \xi+\frac{n(n+2)}{4}\beta^{2}+d_{j}(n)$
 \bd
 \hs{-1cm}\mb{[ i.~e.\ }C_{+}=\alpha=\pm \beta^{2}-C_{-}, \, C_{0}=-(n+1)\beta^{2},\alpha \neq 0\mb{ ]}
 \ed
 \bd
  V_{2}(x)=-\frac{\alpha^{2}}{8\beta^{2}}\cos 2\beta(x-a)+\alpha (n+1)\cos \beta(x-a)-
                                \frac{\beta^{2}}{4}
 \ed
 \bd
 \hs{1cm}E_{j}=\frac{n(n+2)}{4}\beta^{2}-\frac{\alpha^{2}}{8\beta^{2}}\pm \frac{\alpha}{2}+d_{j}(n), \qd
                        j=0,1,\ld ,n \, ,
 \ed
 \bd
 \hs{-1cm}\psi_{j}(x)=\sin \lt \{ (\delta_{k+})\frac{\pi}{2}+\beta \frac{x-a}{2}\rt \}
                              \exp [\frac{\alpha}{\beta^{2}}\sin^{2}(\beta \frac{x-a}{2})]
           \sum_{r=0}^{n}b_{j}^{(r)}\cos^{r}\beta (x-a), \, (k=+,-)
 \ed
 \item $B_{3}=\pm \alpha \xi^{2}-3\beta^{2}\xi \mp \alpha, \, B_{2}=\frac{n(n+4)}{4}\beta^{2}+d_{j}(n)
                                     \mp n\alpha \xi$
 \bd
 \hs{-1cm}\mb{[ i.\ e.\ }C_{+}=\pm \alpha=-C_{-}, \, C_{0}=-(n+2)\beta^{2},\alpha \neq 0\mb{ ]}
 \ed
 \bd
 V_{3}(x)=-\frac{\alpha^{2}}{8\beta^{2}}\cos 2\beta (x-a)\pm \alpha(n+\frac{3}{2})\cos \beta (x-a)
                           -\frac{\beta^{2}}{4}
 \ed
 \bd
 \hs{.5cm}E_{j}=\frac{n(n+4)+3}{4}\beta^{2}-\frac{\alpha^{2}}{8\beta^{2}}+d_{j}(n), \qd j=0,1,\ld ,n \, ,
 \ed
 \bd
 \psi_{j}(x)=\sin \beta(x-a)\exp [\pm\frac{\alpha}{\beta^{2}}\sin^{2}\beta \frac{x-a}{2})]
                \sum_{r=0}^{n}b_{j}^{(r)}\cos^{r}\beta (x-a).
 \ed
 \item $B_{3}=\pm \alpha \xi^{2}-\beta^{2}\xi\mp \alpha, \, B_{2}=\mp n\alpha \xi+\frac{n^{2}}{4}\beta^{2}
                                                                  +d_{j}(n)$
 \bd
 \hs{-1cm}\mb{[ i.\ e.\ }C_{+}=\pm \alpha=-C_{-}, \, C_{0}=-n\beta^{2},\alpha \neq 0\mb{ ]}
 \ed
 \bd
  V_{4}(x)=-\frac{\alpha^{2}}{8\beta^{2}}\cos 2\beta (x-a)\pm \alpha (n+\frac{1}{2})\cos \beta (x-a)
                       -\frac{\beta^{2}}{4}
 \ed
 \bd
 \hs{1cm}E_{j}=\frac{n^{2}-1}{4}\beta^{2}-\frac{\alpha^{2}}{8\beta^{2}}+d_{j}(n), \qd j=0,1,\ld ,n \, ,
 \ed
 \bd
 \psi_{j}(x)=\exp [\pm \frac{\alpha}{\beta^{2}}\sin^{2}\beta \frac{x-a}{2}]
       \:   \sum_{r=0}^{n}b_{j}^{(r)}\cos^{r}\beta (x-a).
 \ed
 \end{enumerate}

 The above potentials are new and appear in the true spirit of QES. It should be noted that the particular
 class corresponding to $n=0$ for $V_{1}(x)$ has been studied for understanding nonaveraged properties of
 disordered systems\ce{tka}. The coefficients $b_{j}^{(r)}, r=1,2,\ld n$ and $d_{j}(n)$ appearing in the 
 Bloch wave functions $\psi_{j}(x)$ and band-edge energies are to be calculated from (9)
 for a given $n$. It is found that number of levels in the algebraic sector is $2|m|$, where $m$ is the
 coefficient of $\alpha \cos \beta (x-a)$ in the potentials.
 \section*{B. Generalized double-well potential} With $u=x-a(a\in \mathbb{R})$, we next adopt for $B_{4}$ 
 the choice $B_{4}=4\gamma^{2}(\xi^{2}-1)$[i.\ e.\ if we set
    $C_{++}=C_{+0}=C_{0-}=0, \, C_{00}=4\gamma^{2}=-C_{--}, \gamma \neq 0$]. $\xi$ turns out to be in the 
 hyperbolic form $\xi=\cosh 2\gamma u$. As before we carry out trials with $B_{3}$ and $B_{2}$ to arrive
 at the following four types of algebraizations:
 \begin{enumerate}
 \item $B_{3}=2\gamma^{2}\eta\xi^{2}+8\gamma^{2}\xi+2\gamma^{2}(\pm 2-\eta), \, 
         B_{2}=-n\gamma^{2}(2\eta\xi+n+2)+d_{j}(n)$ 
 \bd
 \hs{-1cm}\mb{[ i.\ e.\ }C_{+}=2\gamma^{2}\eta=\pm 4\gamma^{2}-C_{-}, \, C_{0}=4\gamma^{2}(n+1), 
                 \eta \neq 0\mb{ ]}
 \ed
 \bd
 V_{1}(x)=\frac{\gamma^{2}\eta^{2}}{8}\cosh 4\gamma (x-a)+2\eta \gamma^{2}(n+1)\cosh 2\gamma (x-a)
                        -\frac{\gamma^{2}\eta^{2}}{8}
 \ed
 \bd
 \hs{1cm}E_{j}=-[(n+1)^{2}\pm \eta]\gamma^{2}+d_{j}(n), \qd j=0,1,\ld ,n \, ,
 \ed
 \bq
 \hs{-1cm}\psi_{j}(x) & = & [(\delta_{k+})\sinh \gamma(x-a)+(\delta_{k-})\cosh \gamma(x-a)] \: 
                                                     \exp \, [\frac{\eta}{4}\cosh 2\gamma(x-a)] \nn \\
  & & \mb{}\times \sum_{r=0}^{n}b_{j}^{(r)}\cosh^{r}2\gamma (x-a), \qd (k=+,-). \nn
 \eq
 \item $B_{3}=2\gamma^{2}(-\eta \xi^{2}+4\xi+\eta \pm 2), \, 
  B_{2}=n\gamma^{2}(2\eta \xi-n-2)+d_{j}(n)$ 
 \bd
 \hs{-1cm}\mb{[ i.\ e.\ }C_{+}=-2\gamma^{2}\eta=\pm 4\gamma^{2}-C_{-}, \,
                 C_{0}=4\gamma^{2}(n+1), \eta \neq 0\mb{ ]}
 \ed
 \bd
 V_{2}(x)=\frac{\gamma^{2}\eta^{2}}{8}\cosh 4\gamma (x-a)-2\eta \gamma^{2} (n+1)\cosh 2\gamma (x-a)
                       -\frac{\gamma^{2}\eta^{2}}{8}
 \ed
 \bd
 \hs{1cm}E_{j}=-[(n+1)^{2}\mp \eta]\gamma^{2}+d_{j}(n), \qd j=0,1,\ld ,n \, ,
 \ed
 \bd
 \hs{-1cm}\psi_{j}(x)=[(\delta_{k+})\sinh \gamma (x-a)+(\delta_{k-})\cosh \gamma (x-a)]\exp [-\frac{\eta}{4}
                      \cosh 2\gamma (x-a)]
 \ed
 \bd
 \hs{1cm}\mb{}\times \sum_{r=0}^{n}b_{j}^{(r)}\cosh^{r}2\gamma (x-a), \qd (k=+,-) \, .
 \ed
 \item $B_{3}=2\gamma^{2}(\mp \eta \xi^{2}+2\xi\pm \eta], , B_{2}=n\gamma^{2}(\pm 2\eta \xi-n)
     +d_{j}(n)$ 
 \bd
 \hs{-1cm}\mb{[ i.\ e.\ }C_{+}=\mp 2\gamma^{2}\eta=-C_{-}, \, C_{0}=4n\gamma^{2}, \eta \neq 0 \mb{ ]}
 \ed
 \bd
 V_{3}(x)=\frac{\gamma^{2}\eta^{2}}{8}\cosh 4\gamma (x-a)\mp 2\eta \gamma^{2}(n+\frac{1}{2})
     \cosh 2\gamma (x-a)-\frac{\gamma^{2}\eta^{2}}{8}
 \ed
 \bd
 \hs{1cm}E_{j}=-n^{2}\gamma^{2}+d_{j}(n), \qd j=0,1,\ld ,n \, ,
 \ed
 \bd
 \psi_{j}(x)=\exp [\mp \frac{\eta}{4}\cosh 2\gamma (x-a)] \, \sum_{r=0}^{n}b_{j}^{(r)}\cosh^{r}
                          2\gamma (x-a).
 \ed
 \item $B_{3}=2\gamma^{2}(\mp \eta \xi^{2}+6\xi\pm \eta), \, B_{2}=n\gamma^{2}(\pm 2\eta \xi-n-4)
     +d_{j}(n)$ 
 \bd
 \hs{-1cm}\mb{[ i.~e.\ }C_{+}=\mp 2\gamma^{2}\eta=-C_{-}, \, C_{0}=4\gamma^{2}(n+2),
                    \eta \neq 0\mb{ ]}
 \ed
 \bd 
 V_{4}(x)=\frac{\gamma^{2}\eta^{2}}{8}\cosh 4\gamma (x-a)\mp 2\eta \gamma^{2}
    (n+\frac{3}{2})\cosh 2\gamma (x-a)-\frac{\gamma^{2}\eta^{2}}{8}
 \ed
 \bd
 \hs{1cm}E_{j}=-(n+2)^{2}\gamma^{2}+d_{j}(n), \qd j=0,1,\ld ,n \, ,
 \ed
 \bd
 \hs{-1cm}\psi_{j}(x)=\sinh 2\gamma (x-a)\exp [\mp \frac{\eta}{4}\cosh 2\gamma (x-a)]
                 \sum_{r=0}^{n}b_{j}^{(r)}\cosh^{r}2\gamma (x-a).
 \ed
 \end{enumerate}

 The potentials $V_{1},V_{2},V_{3},V_{4}$ may be looked upon as hyperbolic counterparts to those of 
 the periodic model. These generalize the bistable potential studied in the context of homonuclear 
 diatomic molecule\ce{raz}. Our potentials are also of interest in spin-boson and spin-spin interacting 
 models where similar hyperbolic forms are known to exist\ce{zas}. As before the coefficients $b_{j}^{(r)}$ 
 in the wave functions and $d_{j}(n)$ in the energies are determined for a given $n$ from (9). The number 
 of analytically known levels is $2|t|$, $t$ being the coefficient of 
 $2\eta \gamma^{2}\cosh 2\gamma (x-a)$.  

 To summarize, we have presented a unified approach of generating ES and QES potentials by exploiting
 the hidden sl(2) symmetry of the Schr\"{o}dinger equation and setting up a master equation. Our scheme
 is especially suitable for generating new types of solvable potentials as we have demonstrated for the QES
 cases.

 \end{document}